\documentclass[12pt,a4paper,oneside]{article}      
\usepackage{amssymb, tabularx, array,booktabs}
\usepackage{amsmath}
\usepackage[authoryear]{natbib} 
\usepackage{graphicx}
\usepackage{algorithmic}
\usepackage{algorithm}
\usepackage{color}
\usepackage[dvipsnames]{xcolor}
\usepackage{verbatim}
\usepackage{rotating}
\usepackage{textcomp}
\usepackage{threeparttable}
\usepackage[font=normalsize]{caption}
\usepackage{verbatimbox}
\usepackage{moreverb}
\usepackage{url}
\usepackage{alltt}
\usepackage{bm}
\usepackage{booktabs}
\usepackage{listings}
\usepackage{enumitem}
\usepackage[a4paper]{geometry}
\usepackage[numbib]{tocbibind}
\usepackage[colorlinks=true,  urlcolor=blue, citecolor=blue, psdextra, pdfencoding=auto]{hyperref}
\usepackage{setspace}
\usepackage[left]{lineno}
\usepackage{ifthen}
\usepackage{soul}

\definecolor{lightviolet}{rgb}{0.97, 0.95, 1.0}

\newcolumntype{C}[1]{>{\centering\arraybackslash}p{#1}}
\newcommand\Fontvii{\fontsize{10}{10.2}\selectfont}
\newcommand\Fontviii{\fontsize{9}{9.2}\selectfont}

\def\thetab{{\bm{\theta}}}
\def\betab{{\bm{\beta}}}
\def\mub{{\bm{\mu}}}

\usepackage{titlesec}

\titleformat*{\section}{\large \scshape}

\def\boxit#1{\vbox{\hrule\hbox{\vrule\kern2pt  \vbox{\kern2pt#1\kern2pt}\kern2pt\vrule}\hrule}}

\begin{document}


\pagenumbering{arabic}



\setstretch{1.25}
\setlength{\linenumbersep}{30pt} 


\title{A fast, flexible simulation framework for Bayesian adaptive
designs - the R package BATSS}

\maketitle

\begin{center}

\author{Dominique-Laurent Couturier$^{1,\dagger}$, Rainer Puhr$^{2}$, Stephane Heritier$^{2}$,\\ Thomas Jaki$^{1,3}$ and Elizabeth G Ryan$^{2}$\\ } 

\vspace{0.5cm}
{\small
$^{\dagger}:$ Corresponding author, \href{Dominique.Couturier@mrc-bsu.cam.ac.uk}{Dominique.Couturier@mrc-bsu.cam.ac.uk} \\
$^{1}:$ MRC Biostatistics Unit, University of Cambridge, UK,\\
$^{2}:$ School of Public Health and Preventive Medicine, Monash University, Australia,\\
$^{3}:$ Faculty of Informatics and Data Science, University of Regensburg, Germany\\
}

\end{center}

\linenumbers

\begin{abstract}
The use of Bayesian adaptive designs for randomised controlled trials
has been hindered by the lack of software readily available to
statisticians. We have developed a new
software package (Bayesian Adaptive Trials Simulator Software - \texttt{BATSS}) for the
statistical software R, which provides a flexible structure for
the fast simulation of Bayesian adaptive designs for clinical trials. We illustrate how the \texttt{BATSS} package can be used
to define and evaluate the operating characteristics of Bayesian adaptive designs for
various different types of primary outcomes (e.g., those that follow a
normal, binary, Poisson, or negative binomial distribution) and can
incorporate the most common types of adaptations: stopping treatments
(or the entire trial) for efficacy or futility, and Bayesian response
adaptive randomisation - based on user-defined adaptation rules. Other
important features of this highly modular package include: the use of (Integrated Nested)
Laplace approximations to compute posterior distributions, parallel
processing on a computer or a cluster, customisability, 
adjustment for covariates and a wide range of available conditional distributions for the response. \\ 

\par \textbf{keywords}: Bayesian adaptive trial; clinical trial simulation; adaptive trial; INLA; R package; Bayesian response adaptive randomisation; multi-arm multistage 
\end{abstract}




\newpage
\section{Introduction}\label{introduction}

Bayesian adaptive designs are increasingly being used to deliver
innovative and complex clinical trials with the aim of providing more efficiency and flexibility (e.g., \citealp{Cellamare2017, effecto2020, denholm2022, gajewski2022}). These designs combine features of Bayesian inference that are attractive for trials,  such as the ability to incorporate prior information like historical controls and to express treatment effects as probability distributions, with the capability to conduct pre-planned analyses on accumulated data at scheduled time points. This allows for alterations to the study design components, such as sample size, randomisation allocation ratios, and the list of treatments available in the trial, based on the results of these analyses.

Multi-arm, multi-stage (MAMS) designs \citep{Royston2003, Royston2011, Magirr2012, Jaki2015} are a type of adaptive clinical
trial design that allow for multiple treatments to be compared within the same trial, often
to a common control treatment, and aim to select the best
treatment(s). This is achieved by performing adaptations such as: early
stopping of treatment arms (removal from the study) for efficacy or
lack-of-benefit (referred to here as ``futility''), early stopping of the trial for efficacy or
futility, and updating the randomisation probabilities based on the
outcome data, to achieve certain experimental goals (known as ``response
adaptive randomisation''). Whilst MAMS designs can be conducted in both Bayesian and frequentist frameworks, we focus here on Bayesian MAMS designs which are naturally suited for adaptations as they can be informed by probability statements on treatment effects (which can be updated at each interim analysis), a feature not accommodated by frequentist MAMS designs. Most of the effort in terms of (open source) software development for MAMS designs has been done in the frequentist framework (e.g., \texttt{nstage} \citep{barthel2009, Blenkinsop2019}, \texttt{MAMS} \citep{jaki2019}, \texttt{rpact}
\citep{rpact}, \texttt{multiarm} \citep{grayling2020a}, \texttt{nstagebin}
\citep{Bratton2014, BabakEtAl2023}, and \texttt{DESMA} \citep{desma}).

\par Bayesian MAMS and other Bayesian adaptive designs  rely on Bayesian inference to adapt trials. At each interim analysis, Bayesian methods combine prior information on model parameters (which might be informed by clinicians, past information, or be vague or non-informative) and the likelihood of parameters given the observed data, resulting in a posterior multivariate probability distribution that summarises the known information about the parameters at a given stage. This distribution allows decision making (such as determining trial or arm successes) to be based on parameter-specific probability statements such as \textit{`there is a 95\% probability that the new treatment is superior to the standard treatment'}. 

\subsection{Monte Carlo Simulations}\label{montecarlosim}

Whilst traditional designs with a single endpoint may rely on (asymptotic) statistical theory to ensure suitable power and control of type I error, such analytical results are typically not available for complex adaptive designs. Consequently, trial operating characteristics (such as power, type I error, and expected sample size) are typically estimated using Monte Carlo simulations in such cases. Among the many research papers and guidelines describing how to perform and evaluate simulation studies (see, e.g., \cite{MorrisSimul2019, WhiteSimul2023, AnkolekarSimul2020, DingSimul2025}), we focus here on those by 
\begin{itemize}[itemsep=-5pt]
    \item the \textit{`International Council for Harmonisation of Technical Requirements for Pharmaceuticals for Human Use'} (ICH), especially `ICH-E20' the new draft guidelines dedicated to adaptive designs [$I_1$: \cite{ICHE20_2025}] and `ICH-E9R1' the draft of the revision of the guidelines dedicated to multiplicity corrections [$I_2$: \cite{ICHE9R1_2017}],
    \item the \textit{`European Medicines Agency'} (EMA), especially the agency initial reflections on adaptive designs [$E_1$: \cite{EMA_AD_2007}] and questions and answers dedicated to complex clinical trials [$E_2$: \cite{EMA_AD_2022}],    
    \item the US \textit{`Food and Drug Administration'} (FDA), especially the agency initial reflections on adaptive Design [$F_1$: \cite{FDA_2019}] and questions and answers dedicated to complex clinical trials [$F_2$: \cite{FDA_2020}],     
  \end{itemize}
 as they provide detailed guidance on expectations to ensure regulatory acceptance and uptake of complex clinical trials, particularly Bayesian ones. For simplicity, to cite these documents, we will use $X^p_d$ to refer to page `p' (subscript) of the document `d' (lower script) of source `X', where `X' can either be `E', `F' or `I', respectively corresponding to the EMA, FDA and ICH in the remaining part of this Section.
The aforementioned guidelines all mention the need for extensive computer simulations to determine important aspects of the design on which the assessment of design proposals is to be made. Such simulations should be ``carefully planned, conducted and reported'' ($I_1^5$) and include, notably, 
\begin{itemize}[itemsep=-5pt]
\item the careful evaluation of the ``chance of producing wrong conclusions'' ($F^4_1$) where the probability of erroneous conclusions should be assessed by estimating the frequentist operating characteristics of type I error by computer-intensive simulations ($F^6_1$, $E^6_1$). When multiple null hypotheses are being tested, the probability of rejecting at least one of them is often controlled ($F^{19}_2$, $E^5_2$), sometimes referred to as the family-wise error rate (FWER). The EMA considers type I error control to be ``a minimum prerequisite for statistical methods to be accepted in the regulatory setting'' ($E^5_1$),
\item a detailed study of other operating characteristics like power and expected, minimum and maximum sample sizes, for example ($F^{11}_2$,$I^{21}_1$),  
\item the optimisation, typically by simulation, of the timing and number of interim analyses ($F^6_1$, $F^{18}_2$),
\item a comparison of the above operating characteristics with the ones obtained when considering alternative designs, including conventional (i.e., non-adaptive) ones ($F^3_1$, $I^{20}_1$).
\item sensitivity analyses to different chosen aspects of the design to ensure operating characteristics are robust to deviations of the assumptions made, notably, 
\begin{itemize}
\item prior choices ($F^4_1$, $I^{25}_1$), especially when estimating the (frequentist-like) type I error and when considering informative priors via the borrowing of external information (past data, expert  opinion). In that case, ``it is also important to evaluate the current trial data with no borrowing'' ($I^{26}_1$).
\item the anticipated timing of interim analyses ($I^{11}_1$), related to, for example, changes to Data Monitoring Committee meeting dates,   
\item  nuisance parameter values ($F^{18}_2$, $I^{21}_1$), for example, 
\begin{itemize}
\item the assumed variance for a Gaussian primary endpoint given the predictors (treatment effects and confounders), 
\item the assumed shape parameter when modelling overdispersed counts with a negative binomial distribution, as well as the average count number in the control group when targeting a given treatment-related fold change, 
\item the probability of success in the control group when modelling a binary endpoint by means of a binomial distribution and targeting a given treatment-related odds ratio as effect size,
\end{itemize}
\item drop-out and missingness levels ($I^{20}_1$),
\item the use of non-binding futility stopping rules when binding ones were planned, as data monitoring committees sometimes opt for flexibility and might consider the futility stopping criteria as a ``guideline that can be deviated from'' ($I^{12}_1$). Of note, in such cases, the FDA would likely consider the trial ``to have failed to provide evidence of efficacy, regardless of the outcome of the final analysis'' ($F^{12}_2$), hence the importance of such sensitivity analyses.
\end{itemize} 
\end{itemize}

\pagebreak
Other relevant aspects of these guidelines include 
\begin{itemize}[itemsep=-5pt]
\item the suggestion, to mitigate the fact that an infinite number of scenarios are compatible with the selected null and alternative hypotheses ($F^{18}_2$, $I^{22}_1$), to determine a limited set of scenarios that adequately represent plausible scenarios ($F^{18}_2$, $I^{13}_1$, $I^{22}_1$). This could be achieved, for example, by relying on medical expert opinion ($F^{19}_2$). The aim would then be to make sure that the maximum type I error across the grid of scenarios does not exceed the target error rate, for example, 2.5\% for a one-sided test ($F^6_1$), and that the power exceeds the target one in all selected cases.  
\item details on the suitable number of Monte Carlo trials per scenario and the way to select seeds per scenario: To achieve a suitable level of precision, it is noted that 100,000 Monte Carlo trials would be sufficient ``in most cases'' when evaluating the type I error, and that 10,000 simulated trials might be enough to evaluate power ($F^{19}_2$, $I^{22}_1$). The use of different seeds per scenario is also recommended ``to avoid [observing] consistently atypical results across scenarios'' ($F^{19}_2$). 
\end{itemize}

\subsection{Existing software}\label{current-software}

\citet{grayling2020} and \citet{meyer2021} provide recent reviews
of software for adaptive designs with the latter focusing on the
software potentially available for simulating MAMS designs,
and both reviews mention some options for Bayesian designs. These reviews found
that simulation of Bayesian adaptive designs generally involved either
constructing bespoke code or using
standalone simulator software (such as FACTS$^{TM}$ \citep{facts}, HECT \citep{thorlund}, or EAST\textsuperscript{\tiny\textregistered}
\citep{east}).

The advantage of these standalone simulators is that they do not require the user
to perform any programming and usually have been optimised to perform the
trial simulations quickly. However, they are often limited in the designs that can be explored (types of endpoints and
models used, adaptations permitted, stopping rules implemented), may
only provide pre-defined operating characteristics, may not allow the user to alter the simulation control parameters (i.e., length of the warm-up/burn-in phase, number of samples, number of chains (for the sampling algorithm)) or provide convergence diagnostics for sampling algorithms, may
require a commercial licence, and/or have graphical user interfaces that
hinder reproducibility.

Script-based software/packages offer several advantages over stand-alone simulators, the main advantages being that they
permit reproducibility, and are generally open-source. However, there is a lack of script-based packages available for simulating 
Bayesian adaptive designs. There are a small number of R
packages for simulating two-arm, Bayesian group sequential designs (\texttt{gsbDesign} \citep{gerber2016}, \texttt{goldilocks} \citep{golilocks})  which allow early
stopping for efficacy or futility, and the \texttt{adaptr} \citep{granholm2022}
package can be used to simulate Bayesian MAMS designs. In
general, these packages can only be used for certain outcomes, are
limited in the types of designs or adaptations that can be implemented
(often with little flexibility in the choice of adaptive rules that can
be used) and may only offer simple models for the outcome. 

For those who do not have access to closed-source simulator software
and/or find existing packages unsuitable for their trial design needs,
simulations will need to be performed using custom-written code. Whilst bespoke code is tailored to meet the needs of a project, it may
only be used for a specific trial design, and may lack the
generalisability to be re-used for future projects \citep{SIMPLE}. Such
code may also lack validation. Additionally, programming
project-specific code may also be an inefficient use of statisticians'
time and lead to delays in starting a clinical trial. 

\par Therefore, the
development of open-source trial simulation software that is reliable, offers good flexibility in the designs and adaptations available, can cope with different types of endpoints
(binary, continuous, count, time-to-event) and is fast is highly desirable.

\subsection{Aim}\label{aim}

The aim of this work is to provide an open-source, script-based, flexible structure (via an R software package) that can be used as a general design tool for the fast simulations of Bayesian adaptive trials and that would enable the efficient evaluation of design characteristics across a wide range of scenarios and  number of simulations, as required by the EMA, FDA, and ICH guidelines mentioned in Section \ref{montecarlosim}. 
\par We will illustrate how our Bayesian Adaptive Trials Simulator Software (\texttt{BATSS}) R package can be used to define and evaluate the operating characteristics of a Bayesian adaptive trial for different types of endpoints given some common adaptations: stopping arms (or the entire trial) for efficacy or futility and response adaptive randomisation. To permit generalisability, we use a modular approach where the user can select the adaptations that they wish to include and specify their own adaptation rules through user-defined R
functions. We provide several examples of such functions to assist the
user. This version of BATSS is suited to trials with slow recruitment and/or those in which endpoints are measured over a short duration with no loss to follow-up. This focus is justified by the fact that long delays in observing the outcome relative to recruitment speed can distort the operating characteristics of the design, and accounting for such delays is therefore left to a future version. We intend to continue the development of BATSS over time to incorporate additional design features.

This package is targeted at statisticians with some experience in adaptive designs, knowledge of Bayesian statistics, who are familiar with R, and who might not have access to commercial clinical trial simulator software or may not want to write their own custom-code to simulate a design. 

This paper details the implementation, methods and functions of the
\texttt{BATSS} R package. In Section 2 we provide a brief overview of the methods used in the \texttt{BATSS} package for approximating the posterior distribution. Section 3 provides an overview of the \texttt{BATSS} framework, functions and workflow. Two case studies, respectively considering a negative-binomial and binomial endpoint, are introduced in Sections 3 and 4, along with an illustration of how these designs would be implemented using the \texttt{BATSS} package and their operating characteristics. We conclude with a discussion on the use of the \texttt{BATSS} package and directions for future extensions of this package.


\section{Laplace approximations for the\\
posterior distribution}\label{fast-approximations-for-the-posterior}

In the Bayesian generalised linear models (GLM) considered in this work, the response vector of $n$ observations, $\mathbf{y} = [y_1,...,y_n]^T$, is assumed to follow a density function depending on parameters $\mub$ and $\phi$
$$ \mathbf{y}\sim f(\mathbf{y}\ |\ \mub,\ \phi)$$ 
where $f(\cdot)$ is a density distribution belonging to the exponential family (or similar), $\mub$ denotes the expectation of $\mathbf{y}$, and $\phi$ is an optional nuisance parameter. The exponential family encompasses a wide range of distributions like the binomial distribution for binary endpoints, the Poisson and negative binomial distribution for count data, and the beta, Gaussian and gamma distributions for continuous endpoints. In GLMs, the response mean is linked to the model predictors through a link function:
$$ \mub = g(\mathbf{X}\betab)$$ 
where $\mathbf{X}$ denotes the ($n \times p$) predictor design matrix, and $\betab$ denotes the regression parameter vector whose interpretation depends on the chosen link function. The expit function $ \mub = \exp(-\mathbf{X}\betab)^{-1}$ is often used for endpoints bounded between 0 and 1 (like binary or beta endpoints), leading to $\betab$ coefficients corresponding to odd ratios. The exponential function $ \mub = \exp(\mathbf{X}\betab)$ is typically used for strictly positive endpoints (like counts), leading to $\betab$ coefficients corresponding to fold changes. The linear model is a special case of GLM in which $f(\cdot)$ corresponds to the Gaussian distribution and $\mub = \mathbf{X}\betab$ (identity link function). 
\par In the frequentist framework, the GLM parameter vector combining the $\betab$ regression parameters and optional nuisance parameter $\phi$, $\thetab = [\betab, \phi]^T$, is estimated by maximising the likelihood function of the parameters given the data. In the Bayesian framework, the parameter vector is assumed to follow a (multivariate) prior distribution $\thetab \sim \pi(\thetab)$. Bayesian inference then asserts that the posterior probability distribution of the fitted model parameters is proportional to the likelihood function multiplied by the prior distribution of the parameters. 
$$\pi(\thetab\ |\ \mathbf{y}, \mathbf{X}) \propto f(\mathbf{y}\ |\ \thetab, \mathbf{X})\ \pi(\thetab)$$

For each iteration of the Monte Carlo
simulations (e.g., 1\ldots M=10,000 trials) that are performed to
understand the operating characteristics of a Bayesian adaptive trial design, the
posterior distribution must be computed (often on multiple occasions within a simulated trial for each interim analysis that is performed). Thus, fast computations of the
posterior distribution are essential. Calculation of the posterior
distribution can often involve high-dimensional integration for which
closed-form expressions do not exist and so numerical techniques are
needed.

Sampling-based methods, such as Markov Chain Monte Carlo (MCMC) methods,
have commonly been used for estimating the posterior distribution and
related quantities for Bayesian inference. These methods require large numbers of MCMC samples which can be
computationally demanding (and sometimes prohibitive), and necessitate both the assessment of convergence of the underlying Markov chain to stationarity as well as the assessment of convergence of Monte Carlo estimators to population quantities \citep{Roy2020, Robert2010}. To circumvent these problems, we use a different approach, which is deterministic and potentially much faster.

\subsection{Integrated Nested Laplace Approximations
(INLA)}\label{integrated-nested-laplace-approximations-inla}

Laplace approximation provides an analytical method to estimate the posterior distribution of parameters in complex models by assuming that the posterior distributions can be approximated by Gaussian densities centered at their mode (or maximum). The calculation of integrals related to the posteriors are therefore greatly simplified, typically leading to computational gains, especially for complex models. R has several implementations of Laplace approximations for Bayesian inference, such as in the TMB (Template Model Builder), LaplacesDemon and mgcv packages, for example.

\newpage
BATSS relies on the Integrated Nested Laplace Approximations (INLA) framework \citep{VanNiekerk2023, BrownEtAl2021, Rue2009} that provides approximate analytical posteriors by assuming that the model follows a specific structure -- the
latent Gaussian model -- and uses sparse matrices \citep{Rue2009}. Latent Gaussian models represent a wide class
of commonly-used statistical models, including generalised linear
(mixed) models, spatio-temporal models, survival and joint longitudinal-survival models. Details about the modelling structure and default priors can be found in \citep{BATSS}, including a dedicated vignette on prior specification (\url{https://batss-stable.github.io/BATSS/articles/web_only/Priors.html}). Users can modify the default priors used by \texttt{R-INLA} via the \texttt{prior} argument in \texttt{batss.glm}.

INLA is implemented in the R software using the \texttt{R-INLA} package \citep{rinla}. INLA has
proved to be popular in many applied areas, including spatial
statistics, ecology, and epidemiology (e.g., \citealp{Alene2022}; \citealp{Pimont2021}) and can potentially be used to perform inference for many types of clinical trial primary outcomes.

In addition to the flexibility provided by the broad range of models that
INLA can accommodate, the computational efficiencies that it offers for performing Bayesian inference were a strong
motivating factor for incorporating INLA into our Bayesian adaptive
trial simulator program. In many situations, INLA methods can provide faster and as accurate inference as MCMC methods \citep{rue2017}. Indeed, \citet{yimer2021} noted, ``The computational time for INLA is in seconds whereas
the computational time for JAGS and Stan is in minutes.'' Parallel processing is also a built-in feature of INLA that further enables computationally efficient Bayesian inference, independent of the specific model that is implemented \citep{muller2019}. In the context of adaptive designs, evidence regarding the computational efficiency of INLA has recently emerged. \citet{hosseini2023} used INLA to ensure the feasibility of the
extensive simulations required to design a Bayesian adaptive trial to
evaluate novel mechanical ventilation strategies in patients with acute
respiratory failure. Speed was particularly an issue for this study based on a hierarchical proportional-odds model. The authors explained that the simulation program would have taken years using R Stan \citep{stan} and MCMC, instead of 20 days with INLA.

\section{The \texttt{BATSS} R package}\label{section-batss}

The aim of the \texttt{BATSS} R package \citep{BATSS} is to provide a user-friendly tool for the fast simulation of Bayesian adaptive trial designs so that a design's operating characteristics can be determined. It is intended that the \texttt{BATSS} package can be used for phase II-IV trials with interim analyses for stopping arms or the trial and alterations to the sample size or allocation ratio. As a first step, we focus our attention on MAMS designs with a common control arm (where we are primarily interested in pairwise comparisons to the control), implemented in the Bayesian framework. 

\texttt{BATSS} is highly customisable: user-defined functions are used to simulate the outcome variable, treatment arm allocations, additional covariates, and also for the adaptation rules. We use a modular structure so that users can select the desired types of adaptations and level of complexity of the design. 
Installation instructions for \texttt{BATSS} can be found in Appendix A.

In this work we define ``intervention'' arms as the experimental treatments that are under investigation (i.e., the non-control arms); ``treatment'' arms refer to all of the arms included in the study, both control and intervention. We define ``active'' treatment arms as a treatment arm (intervention or control) that is still in the trial and has not been dropped. We also use the terms ``adaptive analyses'' and ``interim analyses'' interchangeably. ``Looks'' of the data include interim/adaptive analyses and a final analysis at the maximum sample size (if the trial does not stop early). The ``target parameters'' are the model parameter estimates that correspond to the treatment effects of interest.


\subsection{Case Study 1 - overdispersed count outcome}
Here we consider a case study to illustrate the different components of the \texttt{BATSS} package. This example is inspired by the CHANGE-MS study \citep{CURTIN2016}, which was a multi-arm, phase II randomised controlled trial that investigated the use of temelimab for the treatment of relapsing-remitting multiple sclerosis. This example trial consists of four treatment arms: 3 interventions - low, medium and high doses of temelimab (referred to as arms ``A'', ``B'' and ``C'', respectively) and a control. The outcome is an overdispersed count outcome (cumulative number of new active lesions) that is modelled by a negative binomial distribution (\citealp{CURTIN2016, Hartung2022}), where a lower value indicates a more favourable response. 

The trial has a maximum sample size of 260 patients. We aim to have 90$\%$ power to detect a 60$\%$ reduction (i.e., a  relative risk equal to 0.4)  in the mean cumulative number of lesions in the highest dose intervention arm (from a mean of 4 (control) to 1.6 lesions). We are interested in stopping intervention arms early both for efficacy or futility. To achieve this, we will perform interim analyses starting at 100 patients completing follow-up (across all treatment arms) and then every 40 patients. The trial will run until a decision has been reached for each intervention arm (stopping arms for efficacy or futility) or once the maximum sample size has been reached (default option in \texttt{batss.glm}). For simplicity, we will use equal allocation probabilities (with the potential of dropping intervention arms) and assume a slow recruitment rate.

\subsection{Overview of the \texttt{BATSS} framework}
\begin{figure}[H]
    \centering
   \hspace*{-2cm} \includegraphics[width=\textwidth]{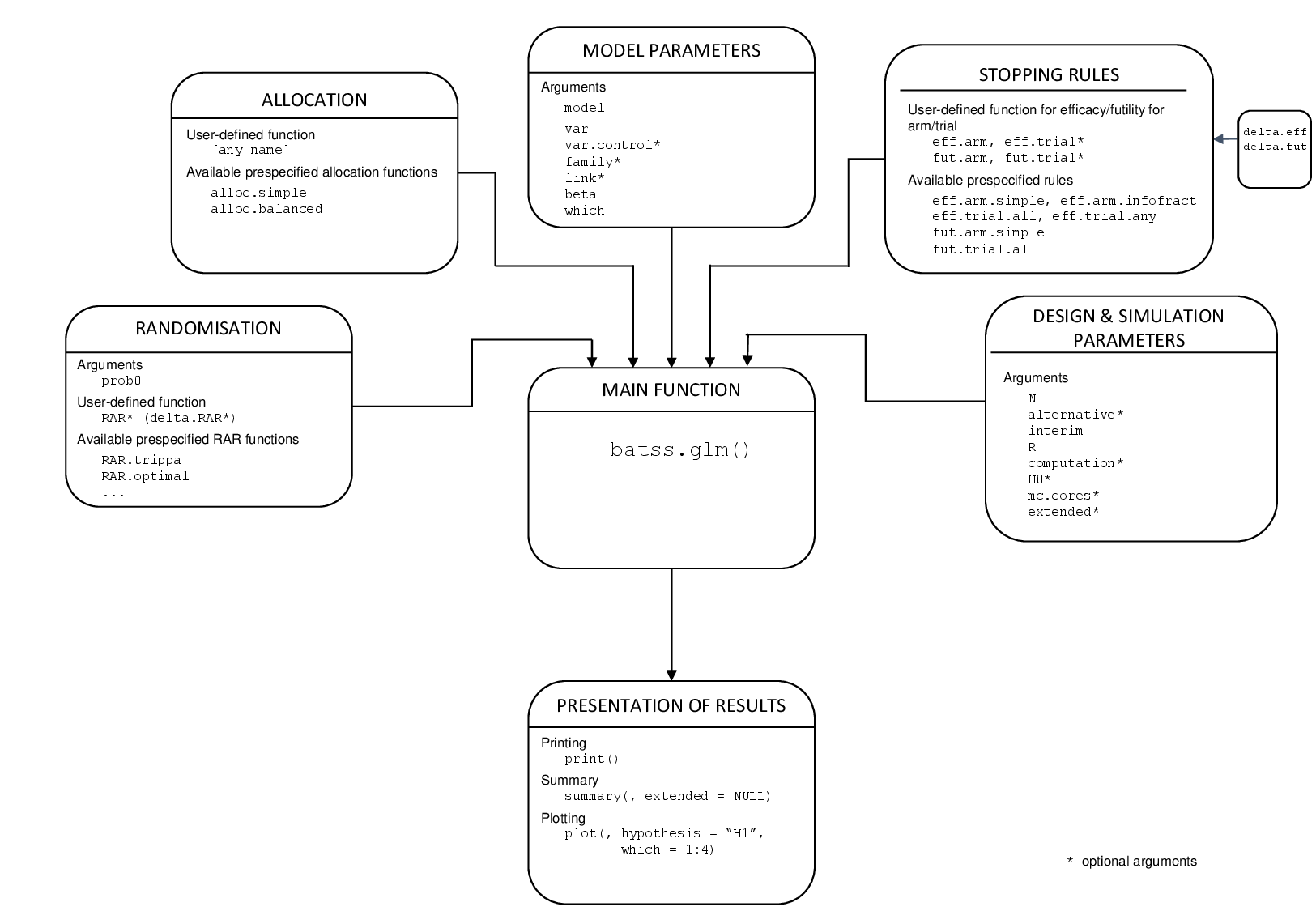}
    \caption{\Fontviii Summary of the inputs and outputs of the \texttt{batss.glm} function, demonstrating its modular structure.}
    \label{fig:bats}
\end{figure}

The overarching function that is used to simulate the trial design is \texttt{batss.glm}, which can be used for designs where the data is modelled by a generalised linear model (GLM). Figure \ref{fig:bats} provides a depiction of the inputs and output generic functions of \texttt{batss.glm}. This function takes a number of arguments that the user can easily alter, and several defaults are provided; these are described in detail Table A1 in Appendix A and in the help function for \texttt{batss.glm}. 

\subsection{\texttt{batss.glm} workflow}

Algorithm \ref{Alg:batsglm}'s pseudocode describes the  process that the \texttt{batss.glm} function performs when simulating a Bayesian adaptive trial design.  A workflow description is also provided in Appendix A.

It is important to note that, if the trial does not stop early, the final analysis is performed at the maximum sample size, using the user-defined efficacy and futility rules for each arm. The maximum sample size is specified for the entire trial (rather than for each arm) and so remaining treatment arms may be enriched with larger sample sizes if some arms are terminated early.

If RAR is selected for the design, it is performed at each adaptive analysis point (after the burn-in period of $m_0$ participants). The user can select whether to implement efficacy or futility stopping from the first adaptive analysis onwards, or to start efficacy or futility stopping at a later analysis timepoint (through the \texttt{delta} arguments of the \texttt{batss.glm} function).

\begin{algorithm}[H]
\Fontvii
\caption{A simplified version of the \texttt{batss.glm} simulation workflow. Relevant arguments of the function appear in {\color{blue}blue}. }
\label{Alg:batsglm}
\begin{algorithmic}[1]
\FOR {$i = 1, ..., {\color{blue}\texttt{R}}$}
\STATE {\color{gray}\#\# Initialisation} 
\STATE Set $m = m_0$ {\color{gray} \ \ \ \# 1st set of patients} {\color{blue}[\texttt{interim=list(recruited=c(m0, ...))}]}
\STATE Set $prob = \color{blue}[\texttt{prob0}] $ {\color{gray}\ \ \ \# initial randomisation probabilities }
\STATE Set $n_{curr} = m$ {\color{gray} \ \ \ \#  current sample size}
\WHILE{Any intervention arms are active and $n_{curr} \le {\color{blue}\texttt{N}}$}
\STATE Simulate $m$ sets of predictors {\color{blue}[\texttt{var}, \texttt{var.control}]} 
\STATE Simulate $m$ responses {\color{blue}[\texttt{model}, \texttt{var}, \texttt{var.control}, \texttt{link}]}
\STATE Fit model to the data using \texttt{INLA} {\color{blue}[\texttt{model}, \texttt{family}, \texttt{link}]}
\STATE Calculate target parameter posterior probabilities using the $\delta$s\\ {\color{blue}[\texttt{which}, \texttt{delta.eff}, \texttt{delta.fut}, \texttt{delta.RAR}]}
\FOR {$k$ in  no. of active intervention arms}
\STATE Assess efficacy stopping rules\ \ {\color{blue}[\texttt{eff.arm}, \texttt{eff.arm.control}]}
\STATE Assess futility stopping rules \ \ {\color{blue}[\texttt{fut.arm}, \texttt{fut.arm.control}]}
\ENDFOR
\STATE Assess trial stopping rules \ \ {\color{blue}[\texttt{eff.trial}, \texttt{fut.trial}]}
\IF{All intervention arms inactive or trial stopping rules have been met}
\STATE BREAK
\ENDIF
\IF{RAR $\neq$ NULL}
\STATE define randomisation probabilities \ \ 
{\color{blue}[\texttt{RAR}, \texttt{RAR.control}, \texttt{delta.RAR}]}
\ENDIF
\STATE Update $m$, $n_{curr}$, and $prob$
\ENDWHILE
\ENDFOR
\end{algorithmic}
\end{algorithm}

\subsection{The \texttt{batss.glm} function - syntax} 

Case Study 1 may be simulated using the following syntax:

\begin{lstlisting}[keywordstyle=\color{black}] 
sim = (*@\textcolor{blue}{batss.glm}@*)(
                 model              = y~treatment,
                 var                = list(y = rnbinom,          
                                           treatment = alloc.balanced),
                 var.control        = list(y = list(size = 1/2)), # shape
                 family             = "nbinomial",
                 link               = "log",
                 beta               = log(c(4, 1, 1, 0.4)),
                 which              = c(2:4),
                 alternative        = c("less"),
                 R                  = 10000,
                 N                  = 260,
                 interim            = list(recruited=c(100,140,180,220)),
                 prob0              = c(control = 1, A = 1, B = 1, C = 1),
                 delta.eff          = 0, 
                 eff.arm            = eff.arm.infofract,
                 eff.arm.control    = list(b.eff = 0.009, p.eff = 3),
                 delta.fut          = log(0.8), 
                 fut.arm            = fut.arm.simple,
                 fut.arm.control    = list(b.fut = 0.2025),
                 RAR                = NULL,
                 RAR.control        = NULL,
                 H0                 = FALSE,
                 computation        = "parallel",
                 mc.cores           = detectCores()-1,
                 extended           = 1)    
           
\end{lstlisting}

Here we use some pre-defined functions for the treatment allocations (\texttt{alloc.balanced}), efficacy (\texttt{eff.arm.infofract}) and futility (\texttt{fut.arm.simple}) stopping.

\begin{figure}
    \centering
    \includegraphics[width=8cm]{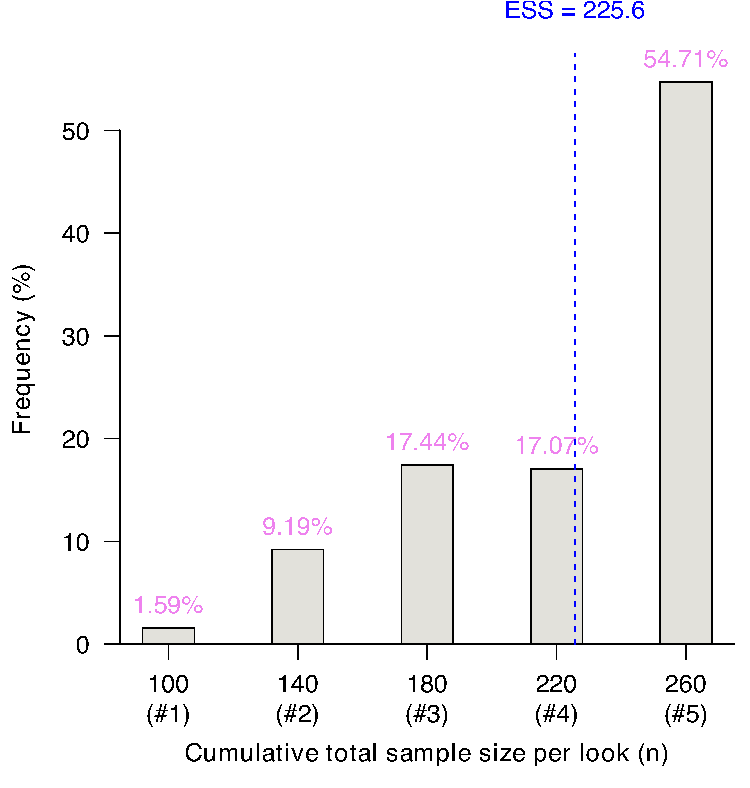}
    \caption{\Fontviii Barplot of the probability of stopping (y-axis) at each look (x-axis) as observed in the simulated samples. The vertical blue dashed line shows the estimate of the expected sample size (ESS).}
   \label{fig:NBplot2}
\end{figure}

\begin{figure}
    \centering
    \includegraphics[width=\textwidth]{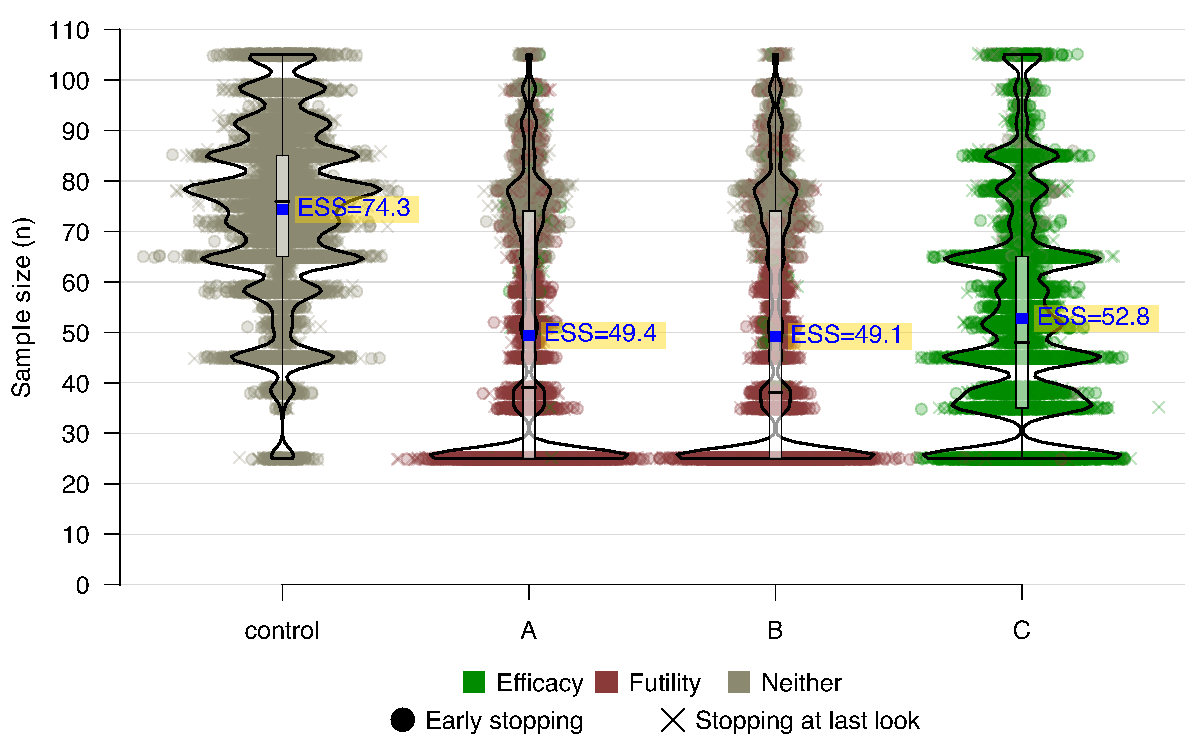}
    \caption{\Fontviii Violin plot (and boxplot) of the simulated sample sizes per treatment arm for each trial and corresponding expected sample sizes. For target parameters, these points are colour-coded and shaped by the decision and timing of when the arm stopped.}
   \label{fig:NBplot3}
\end{figure}

\begin{figure}
    \centering
    \includegraphics[width=\textwidth]{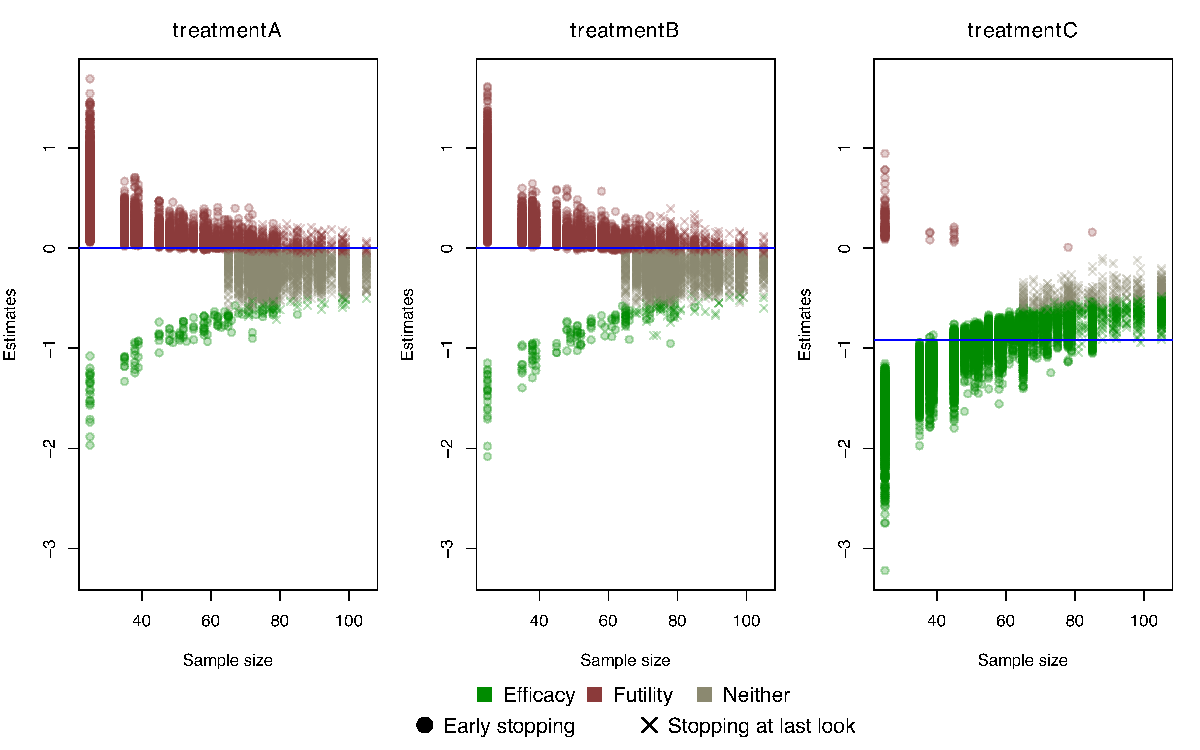}
    \caption{\Fontviii Scatterplot of target parameter estimates (posterior mean) versus sample size for each intervention arm (separate plots) at the time of stopping the arm, along with the decision at the time of stopping for the simulated scenario for Case Study 1 across the 10,000 simulated trials. The timing of the stop is also provided (“Early stopping” = stopping at an interim analysis).The horizontal blue lines show the assumed target parameter values.}
   \label{fig:NBplot4}
\end{figure}

The operating characteristics for the above simulated scenario can be obtained using:
\begin{lstlisting}
    (*@\textcolor{blue}{summary}@*)(sim, extended = 1) 
\end{lstlisting}

Figure \ref{fig:NBplot2} shows the probability of stopping the trial at each look under the assumed parameter values as well as the expected trial sample size. Figure \ref{fig:NBplot3} displays the sample size of each simulated trial per group with, for the target parameters, points colour-coded and shaped according to the decision and timing when the trial stopped. Figure \ref{fig:NBplot4} shows the parameter estimates (posterior mean) against the sample size for each arm and the decision at the time of stopping the arm for the simulated design. These plots  can be obtained using: 
\begin{lstlisting}
    (*@\textcolor{blue}{plot}@*)(sim, which = 2:4)
\end{lstlisting}
where \texttt{which=1} produces the boxplot of the total and per group sample sizes of the simulated trials. 

A fuller description of Case Study 1, along with the full code to simulate the example, output and interpretation of the output can be found in Appendix B. 

We now briefly describe the arguments of the \texttt{batss.glm} function (and provide a more detailed description in Table A1 in Appendix A). A vignette is also available in \cite{BATSS}.

\newpage
\subsection{Defining the model - types of outcomes, data simulation and model fitting}

\texttt{R-INLA} provides inference for many types of GLMs and allows for several different families for each data type, such as: 

\begin{itemize}[itemsep=-5pt]
    \item Binary data: binomial, beta-binomial,
    \item Count data: Poisson, zero-inflated Poisson, negative binomial,
    \item Continuous data: Gaussian, Weibull, exponential, lognormal, gamma,
    \item Bounded data: beta.
\end{itemize}

This offers additional flexibility over many existing adaptive design simulation packages, which are often limited in the distributions that they support. The above list is not exhaustive. Users can check the full list of distribution families available in \texttt{R-INLA} by typing \texttt{names(INLA::inla.models()\$likelihood)} in R after loading the \texttt{INLA} package.

Our framework can handle general cases in which the response is a function of the treatment arm allocation and additional predictors, with or without interaction terms, and in which the target parameters can include both the main and/or interaction parameters. However, for illustrative purposes, we consider simpler models in this paper and work under the assumption that the user-specified model is a GLM with a linear predictor ($g(\mu)$) of the form 
\begin{equation}\label{eq:glm}
    g(\mu) = \mathbf{X}\bm{\beta} = \beta_0 + \sum_{k=1}^{K}\, \beta_k \, I_k
    \end{equation}
where $\mathbf{X}$ denotes the model design matrix and $\bm{\beta}=[\beta_0,...,\beta_K]^T$ is the corresponding parameter vector, and $I_k$ is an indicator for treatment arm $k$ (of which there are $K$ interventions to compare to control; $K+1$ arms in total included in the study).

In the \texttt{batss.glm} function, the user specifies the family type (\texttt{family = }) and link function (\texttt{link = }) that are to be used when fitting the model (these must correspond to the options provided in \texttt{R-INLA}). The user must also specify a model object (\texttt{model = }) indicating both how to simulate and fit the response (i.e., the same model is used for both purposes). Predictor variables (such as the treatment allocation (factor)) are simulated according to user-specified functions defined in the \texttt{var} and \texttt{var.control} arguments and the user-specified model is used to generate the linear predictor value ($\mathbf{X}\bm{\beta}$). This is then used to simulate the response, using the inverse link function and the user-specified outcome distribution and parameters (defined  in the \texttt{var} and \texttt{var.control} arguments). 

Additional predictors can be added to the model by including their variable name in the \texttt{model} formula, as well as their data generating process and related parameter values in the \texttt{var} and \texttt{var.control} arguments. Note that it is possible to generate dependent predictors by specifying a multivariate data generating process in the \texttt{var} argument.

For Case Study 1, a dispersion (shape) parameter with a value of 0.5 was assumed using the following parameterisation for the variance:
\begin{equation}\label{eq:1}
 Var(Y)= \mu + \mu^2/\phi
\end{equation}

(i.e., $\phi=1/2$). This is a nuisance parameter that has to specified and this is done via the command \texttt{var.control= list(y = list(size = 1/2))}. 

The user can specify the assumed true parameter values (corresponding to those in equation \ref{eq:glm}) via the \texttt{beta} argument (on the link function/linear predictor scale). The interpretation of the elements of this vector matches the one we would have in a fit obtained via the \texttt{lm} or \texttt{glm} R functions, and therefore depends on the way predictors are generated and contrasts of factors are defined. For Case Study 1, using the model specified in (\ref{eq:glm}), assuming treatment or dummy contrasts \citep{RBookLinearModel} are used for the factor `treatment' with `control' as the reference group and the use of the log link, then the first value of \texttt{beta} corresponds to the mean response in the control treatment (the intercept) on the log scale, and the second, third and fourth elements correspond to the logarithm of the relative risks for treatments A, B and C compared to control. If we assume a mean count of 4, 4, 4 and 1.6 for the control and groups A, B and C, respectively (on the original scale), we would then have the model $g(\mu) = log[E(Y)]=log(4)+ log(1)\,I_A +log(1)\,I_B +log(0.4)\, I_C$ (where $I_A$, $I_B$ and $I_C$ are indicator variables for treatment arms A, B and C, respectively) and $\bm\beta=[log(4), log(1),log(1),log(0.4)]^T$. The user then specifies which of those parameters correspond to the parameters of interest (known as the ``target'' parameters), via the \texttt{which} argument. In our model, we are interested in the incremental effects of treatments A, B and C with respect to the control so that the target parameters are the second, third and fourth elements of \texttt{beta} and \texttt{which=c(2:4)}.

\subsection{Design specification - interim analyses and adaptations}

The direction of the alternative hypothesis is specified via the \texttt{alternative} argument for each target parameter  (one-sided hypothesis testing), where \texttt{alternative="less"} in this instance as a lower value indicates a more favourable response. The maximum sample size of 260 is provided via the \texttt{N} argument. 

The user provides the timing of the interim analyses in the \texttt{interim} argument (where \texttt{interim=NA} leads to the use of a fixed design). Currently, \texttt{batss.glm} assumes that the outcome data is immediately available and that interim analyses can only be performed based on the number recruited (\texttt{interim = list(recruited = )}). Thus, here, interim analyses based on the number recruited is equivalent to interim analyses based on the number with complete follow-up, i.e., the same amount of information will be available under both monitoring schedules. The interim analyses can be equally-spaced or unequally spaced. 

The \texttt{batss.glm} function supports the following types of adaptations: stopping arms for efficacy (also known as ``graduation'' of arms), stopping (dropping) arms for futility, stopping the entire trial early to declare one or more interventions superior to control, stopping the entire trial early due to statistical futility (e.g., lack-of-benefit), sample size re-estimation, and response-adaptive randomisation (RAR). Fixed allocation ratios can also be used instead of RAR (\texttt{RAR = NULL}). More generally, \texttt{batss.glm} can also be used to simulate fully fixed (non-adaptive) designs by setting \texttt{delta.eff} and \texttt{delta.fut} to \texttt{NA} at a single, dummy interim analysis and \texttt{RAR} to \texttt{NULL}. Code examples for fixed designs are provided in the \texttt{BATSS} vignettes \citep{BATSS} and in Appendices~B and~C.

As a design's operating characteristics are strongly influenced by the chosen efficacy and futility stopping rules as well as the chosen method to set the group allocation probabilities, flexibility in defining these adaptations is of crucial importance. \texttt{BATSS} therefore allows these rules to be user-defined, thus allowing great flexibility.

The efficacy and futility stopping rules (for the intervention arms) are provided via user-defined functions in the \texttt{eff.arm} and \texttt{fut.arm} arguments, respectively (with input parameters provided in  \texttt{eff.arm.control} and \texttt{fut.arm.control}). Efficacy and futility stopping rules of the trial at a given interim analysis are respectively indicated via the \texttt{eff.trial} and \texttt{fut.trial} arguments that take the target parameters results as input. This, for example, allows the trial to stop once one intervention arm has been found efficacious. Several example functions are provided in the \texttt{BATSS} vignette \citep{BATSS}. We defer their detailed explanation to the case studies (Section 4, and Appendices B and C).

Initial randomisation probabilities/allocation ratios, prior to RAR implementation (i.e., during the ``burn-in'' period before the first interim analysis), are provided in \texttt{prob0}. These probabilities are used throughout the trial if RAR is not implemented. Here we chose equal randomisation throughout the trial, therefore \texttt{prob0 = c(control = 1, A = 1, B = 1, C = 1)} (note that one could also set \texttt{prob0 = c(control = 1/4, A = 1/4, B = 1/4, C = 1/4)}). These probabilities will be re-scaled if a treatment is dropped.

The user-defined RAR rules are provided via the \texttt{RAR} argument (with input parameters for the RAR function given in \texttt{RAR.control}). The user-defined functions permit restrictions of RAR, such as allocation rules with fixed proportions to control or minimum or maximum allocation probabilities to be implemented. This will be illustrated Case Study 2 and several examples are provided in the \texttt{BATSS} vignette.

These different adaptations typically rely on the posterior probability of being greater or smaller than the clinically important difference indicated in the delta arguments -- respectively \texttt{delta.eff} and \texttt{delta.fut} for the efficacy and futility arm stopping rules, and \texttt{delta.RAR} for the RAR probabilities  -- which is estimated at each look for each target parameter.  

We provide some examples of functions for the adaptations in Section \ref{binary-outcome} and Appendices B and C, and describe the required inputs for these functions in Table \ref{tab:table_ingredients} and Table A2 in Appendix A. When relying on self-defined functions (for treatment allocation, efficacy and futility stopping rules, and RAR), we recommend that users test them outside of \texttt{batss.glm} before using them in a simulation, both to identify potential errors and to verify that the functions behave as intended. Examples of such tests are provided in the \texttt{BATSS} vignettes \citep{BATSS}.

\subsection{\texttt{BATSS} ingredients}

The \texttt{batss.glm} function calculates a number of ``ingredients'' or derived quantities at each interim analysis which can be used as inputs to the user-defined functions for the adaptations. These are summarised in Table \ref{tab:table_ingredients} (full description given in Table A2 in  Appendix A, and in the vignette \citep{BATSS}).

\begin{table}[]
    \centering
    \Fontviii
\begin{tabularx}{5in}{p{1in}p{3.5in}}\hline
        \textbf{Quantity} & \textbf{Description} \\
        \cmidrule(r){1-1}
        \cmidrule(r){2-2}
         active & A logical vector indicating which treatment arms are still active/remaining in the trial. \\
         posterior & Posterior probability of efficacy (futility) of an ``active'' intervention, incorporating the deltas (clinically important effects). \\
        n & A vector providing the number of participants that have been recruited in each treatment arm; includes control and intervention arms (both active and dropped arms).\\
        ref & A logical vector indicating which treatment arm is the reference (control). \\
        prob & Randomisation probabilities (for the active arms). The output from the RAR function (unnormalised). If RAR=NULL in \texttt{batss.glm} then \texttt{prob=prob0}. \\
        m & number of subjects recruited since the last interim analysis\\\hline
        \end{tabularx}     
    \caption{Description of the common quantities that \texttt{batss.glm} calculates at each interim analysis that can be used as inputs to the user-defined functions for  adaptations}
    \label{tab:table_ingredients}
\end{table}

\subsection{Parallel and cluster computing}

In \texttt{BATSS}, the number of simulated trials is provided by means of the argument \texttt{R}, either with a scalar denoting the number of trials (in that case seeds 1 to R will be used, e.g., 1 to 10,000 for our example above) or with a vector of seeds to use to generate each trial's data (in that case the number of trials will equal the length of that vector).

The \texttt{BATSS} package takes advantage of computing tools for embarrassingly parallel workloads which allows the user to split the burden of the independent Monte Carlo trials between multiple cores on single computers or multiple cores/Central Processing Units (CPUs) of different computers/clusters:
\begin{itemize}
    \item When performing computation on the same computer, the user can use parallelisation tools of the \texttt{parallel} package by setting the argument \texttt{computation} to `parallel' and by specifying the number of child processes -- Monte Carlo trials in this context -- to be run in parallel via the argument \texttt{mc.cores}.
   \item To split the computational burden between different computers and/or the CPUs of one or several clusters, the user can split the planned list of seeds into non-overlapping sets. Each set will then be assigned to an available resource (like a CPU on a cluster), either by hand or more efficiently via a script (like an sbatch on clusters using Simple Linux Utility for Resource Management (SLURM) as the job manager), by a call of the \texttt{batss.glm} function using the set of seeds of interest through the \texttt{R} argument and saving the results of each trial (setting argument \texttt{extended} $>$ 0). The \texttt{batss.combine} function, taking as input the location of the R data files containing the outputs corresponding to each set of seeds, will then be used to aggregate results. (Note that  \texttt{batss.combine} checks if the sets of seed overlap or if the calls to \texttt{batss.glm} were different).   
\end{itemize}

We provide some exemplar code in a vignette demonstrating how to run the \texttt{batss.glm} function on a cluster and how to use the \texttt{batss.combine} function (see \cite{BATSS}) .

\subsection{Output}

The output generated by \texttt{batss.glm} is an S3 object of class \textit{batss} (with \texttt{type = glm}), with available  \texttt{print}, \texttt{summary} and \texttt{plot} `methods'. A number of different operating characteristics are provided in the output generated by \texttt{batss.glm}. 

The summary method (obtained by using the \texttt{summary} function on the output of \texttt{batss.glm}) first prints a summary of the design and inputs to the \texttt{batss.glm} function. Then, in its ``short'' variant (obtained by opting for \texttt{extended = 0}), it indicates the probability of declaring efficacy or futility for each intervention arm, for at least one intervention arm, as well as for all intervention arms for each hypothesis of interest (i.e., under the null and/or alternative hypotheses if argument \texttt{H0=TRUE}, for example). Note that, as a diagnostic tool, the probability of simultaneously meeting both the efficacy and futility criteria is also provided if this highly undesirable situation occurred for any arm in any simulated trial. In its ``long'' variant (obtained using the \texttt{summary} function with \texttt{extended} set to \texttt{1} or \texttt{2}), the method additionally provides: a breakdown of when the efficacy or futility stopping occurred (early = at an interim analysis; last = at the maximum sample size); summaries of the sample sizes for each arm and overall (mean, standard deviation, median and 10th and 90th percentiles); and a summary of the different scenarios that occurred in terms of the decisions for each intervention arm (no decision, efficacy decision, futility decision, both efficacy and futility criteria met (this is only displayed for simulated designs where this undesirable behaviour has occurred)) and how frequently those different combinations occurred across the Monte Carlo trials (and what proportion of those had early decisions made where the trial stopped at an interim analysis). 

By using the \texttt{extended} argument in the \texttt{batss.glm} function, users can specify what they want to be stored in the output: \texttt{0} only saves the overall summary operating characteristics, \texttt{1} adds the trial results with a breakdown by look/analysis point to them, and \texttt{2} includes all previous results plus the R simulated datasets. 

We recommend that users first inspect the results of a few individual simulated trials (available when \texttt{extended = 1} or \texttt{extended = 2}) before examining the summary operating characteristics. This allows users to verify that the design is behaving as intended -- for example, that the stopping rules and response-adaptive randomisation are triggering correctly at each interim analysis -- and can also serve as a useful tool for communicating the design to stakeholders. Narrative examples of such inspections are provided in Appendices~B (Section 2) and~C (Section 1.3). We also recommend that users simulate a corresponding fixed (non-adaptive) design early in the design process. This serves two purposes: it provides a simpler baseline against which to verify that the code is working correctly, and it allows the operating characteristics of the adaptive design (e.g., power, expected sample size, probability of early stopping) to be compared against those of the fixed design, thereby quantifying the benefits of the contemplated adaptations. Examples of such comparisons are provided in Appendices~B and~C.

Plots of the posterior mean of the target parameter (i.e., treatment effect) estimates versus the sample size at the last look performed for that Monte Carlo trial, along with the decision (efficacy, futility, none, both), for each intervention arm, can be obtained using the \texttt{plot} function with the \texttt{which = 4} option (demonstrated in Figure \ref{fig:NBplot4}). Plots of the distributions of the sample sizes per group (across the Monte Carlo trials) can also be obtained using the \texttt{which = 3} option in the \texttt{plot} function (demonstrated in Figure \ref{fig:NBplot3}), as well as using the \texttt{which = 1} option.

We present a second case study in Section \ref{binary-outcome} to further illustrate the use of the \texttt{batss.glm} function and how to summarise its output.

\subsection{Validation of BATSS}

Several complementary approaches were used to verify the correctness 
of the \texttt{BATSS} software throughout its approximately five-year 
development period.
\begin{itemize}[itemsep=-2pt]
    \item The entire codebase was reviewed line-by-line by multiple co-authors (DLC, RP, and EGR), and simulation outputs were collectively examined to ensure that designs behaved as expected under known scenarios.  Version~1.1.1 of the source code was additionally reviewed using the large language model Claude (Opus 4.6; \citep{anthropic2025claude}),
    \item To validate the simulation workflow, we compared the frequentist operating characteristics produced by a frequentist counterpart of \texttt{BATSS} (which replaces posterior-based decisions with classical test statistics and group-sequential boundaries) against results from the established \texttt{MAMS} R package \citep{jaki2019}. Agreement was verified across several MAMS design configurations, including designs with simultaneous and separate stopping rules for treatment arms as well as drop-the-loser designs. 
    \item The package was stress-tested under extreme parameter settings (e.g., very small or very large treatment effects, near-zero event rates, high overdispersion) to assess numerical stability.
    \item The Bayesian inference engine used by \texttt{BATSS} (\texttt{R-INLA}) has itself been extensively validated against MCMC methods for the class of generalised linear models relevant to \texttt{BATSS} \citep{Held2010, rue2017}.
\end{itemize}


\section{Case Study 2 - binary outcome}\label{binary-outcome}

Here we demonstrate how to use the \texttt{batss.glm} function to simulate a Bayesian MAMS trial design with a binary outcome. This example is motivated by the Personalised Immunotherapy Platform (PIP) trial \citep{PIP}, which is a phase II randomised controlled trial that is currently under development that will compare immunotherapy combinations for advanced melanoma patients.

This study aims to assess the effectiveness of five novel drug combinations (named here arms ``B-F'') compared to a control arm (standard of care; arm ``A'') in treatment-naive patients that are predicted to be unlikely to respond to standard first line immunotherapy treatments. The primary outcome is the objective response rate (ORR) according to the Response Evaluation Criteria in Solid Tumors (RECIST) (partial or complete response vs stable disease or progressive disease) at 6 months post randomisation. We assume that this study is slow-recruiting relative to the time to obtain the primary endpoint. Here we are interested in the pairwise comparisons of each intervention arm with the control to demonstrate whether the novel drugs are superior to the standard of care. Head-to-head comparisons of the novel drugs is not of primary interest. We assume an ORR of 40$\%$ in the control arm and expect a poorly-performing novel drug to lead to no more than a 10$\%$ absolute improvement when compared to the control, and expect to observe a minimum absolute improvement in ORR of 20$\%$ for an effective novel drug.  

Interim analyses will be performed once 60 participants have completed follow-up, and then every 12 participants thereafter, up to a maximum sample size of $N=216$ (leading to a total of 14 analyses). A burn-in period of 60 participants will be used during which equal randomisation will be implemented (1:1:1:1:1:1). After this, Bayesian RAR will begin. The randomisation probabilities are calculated using a variation of \cite{trippa2012} and \cite{Cellamare2017} where the allocation probabilities at look $j$ are proportional to:
\begin{itemize}
\item  control (reference) arm: 
\begin{equation}\label{eq:rar_cont}
   \frac{ \exp(\text{max}(n_{1,j}, n_{2,j}, ..., n_{5,j})-n_{0,j})^{\nu}}{K_j}
\end{equation}

 with $(n_{1,j}, n_{2,j}, ..., n_{5,j})$ being a vector of the number of participants recruited to each intervention arm (arms B-F) at look $j$, $n_{0,j}$ is the number of participants recruited to the control arm at look $j$, $K_j$ = number of active intervention arms ($K_j +1$ is the number of active arms including control) at look $j$ (i.e., number of arms still in the study at that point in time), $\nu$ is a tuning parameter.
\item intervention arms (where $k>0$) : 
\begin{equation}\label{eq:rar_int}
\frac{P(\beta_{k} > \delta_{r} |\ \mathbf{y}, \mathbf{X})^{h}}{\sum^{K_j}_{l=1}P(\beta_{l} > \delta_{r}  |\ \mathbf{y}, \mathbf{X})^{h}}
\end{equation}
with $h=\gamma(\sum^{K}_{k=0} n_{k,j}/N)^{\eta}$, where $n_{k,j}$ is the number of participants enrolled in arm $k$ at look $j$ ($K$ is the total number of intervention arms initially included in the study, not including the control; $k=0$ is control), and where $\mathbf{y}$, $\mathbf{X}$ respectively correspond to the vector of responses and model design matrix. Here $\beta_{k}$ is the treatment effect for intervention arm $k$ (log-odds scale) and $\gamma$ and $\eta$ are tuning parameters (see \citet{Gotmaker}). $\delta_{r}$ is the (RAR-related) clinically meaningful treatment effect value, and for this example we set $\delta_{r}=0$.
\end{itemize}

These RAR functions (equations \ref{eq:rar_cont} and \ref{eq:rar_int}) can be written in R as:

\begin{lstlisting}
prob.trippa = (*@\textcolor{blue}{function}@*)((*@\textcolor{orange}{posterior, n, N, ref, active}@*), (*@\textcolor{red}{gamma,eta,nu}@*)){
  g = sum((*@\textcolor{orange}{active}@*)) # (K_j+1 in the above)
  h = (*@\textcolor{red}{gamma}@*)*(sum((*@\textcolor{orange}{n}@*))/(*@\textcolor{orange}{N}@*))^(*@\textcolor{red}{eta}@*)
  # vector of allocation probabilities:
  prob = rep(NA,g) 
  # reference/control arm allocation - Equation 3:
  prob[1] = (exp(max((*@\textcolor{orange}{n}@*)[!ref])-(*@\textcolor{orange}{n}@*)[ref])^(*@\textcolor{red}{nu}@*))/(g-1)
  # targets/interventions (that haven't been dropped) - Equation 4:
  prob[2:g] = ((*@\textcolor{orange}{posterior}@*)^h)/(sum((*@\textcolor{orange}{posterior}@*)^h))
  prob   
}
\end{lstlisting}


where the arguments in \textcolor{orange}{orange} are \texttt{BATSS} ingredients (Table \ref{tab:table_ingredients}), and those in \textcolor{red}{red} are items from the respective user-provided `\texttt{.control}' arguments in the \texttt{batss.glm} function. This same convention applies throughout this section.

We would then input these probabilities into a treatment allocation function, such as:
\begin{lstlisting}
# specify group allocation function
group.fun = (*@\textcolor{blue}{function}@*)((*@\textcolor{orange}{m, prob}@*)){
  (*@\textcolor{orange}{prob}@*) = abs((*@\textcolor{orange}{prob}@*))/sum(abs((*@\textcolor{orange}{prob}@*))) # (normalise probabilities)
  factor(sample(names((*@\textcolor{orange}{prob}@*)), (*@\textcolor{orange}{m}@*), replace=TRUE, 
         prob = (*@\textcolor{orange}{prob}@*)), levels=names((*@\textcolor{orange}{prob}@*)))
}
\end{lstlisting}

Using these RAR functions (equations \ref{eq:rar_cont} and \ref{eq:rar_int}), the allocations to the control arm are protected so that sufficient power can be obtained for the comparisons of interest.

Stopping arms for efficacy is not permitted at the interim analyses in this case study as the main objective of the adaptations is to drop poorly performing arms. 

An intervention arm $k$ may be stopped for futility if there is a low posterior probability of observing at least a 10$\%$ absolute increase (improvement) in the proportion of responders, i.e.,

\begin{equation}\label{eq:pip_fut1}P(\beta_{k} > \delta_f |\ \mathbf{y}, \mathbf{X}) \ \ < \ \ b_f
\end{equation}

where $\beta_{k}$ is the treatment effect for intervention $k$ (on the log-odds scale), and $\delta_f$ is the minimum clinically important treatment effect (on the log-odds scale), i.e.,
\[ \delta_f = log\left(\frac{(\pi_0+0.1)/(1-(\pi_0+0.1))}{\pi_0/(1-\pi_0)}\right) \]
($\pi_0$ is the control arm ORR and $\pi_0 + 0.1$ is the minimum desired ORR in an intervention arm). 

We begin by using a fixed cut-off of the futility boundary of 0.1, i.e., $b_f=0.1$ (Case Study 2.1). This futility stopping rule (equation \ref{eq:pip_fut1}) can be written in R as:
\begin{lstlisting}
futility1.arm.fun = (*@\textcolor{blue}{function}@*)((*@\textcolor{orange}{posterior}@*), (*@\textcolor{red}{b.fut}@*)){
  (*@\textcolor{orange}{posterior}@*) <  (*@\textcolor{red}{b.fut}@*)
}
\end{lstlisting}

We will also investigate a stopping rule where the futility boundary increases as a function of the existing number of participants that have been randomised (Case Study 2.2): 

\begin{equation}\label{eq:pip_fut2}P(\beta_{k} > \delta_f |\ \mathbf{y}, \mathbf{X})  \ \ < \ \ b_f \left(\frac{\sum^{K}_{k=0} n_{k,j}}{N}\right)^{p_f},
\end{equation}

where $b_f$ is the futility (lack-of-benefit) boundary at the maximum sample size, $\sum^{K}_{k=0} n_{k,j}$ is the total number of participants enrolled at look $j$, and $p_f$ is a tuning parameter that dictates the shape of the function \citep{Gotmaker}. That is, equation 
\ref{eq:pip_fut2} is an increasing power function of the fraction of information available in the trial at that stage.  Here we use an increasing cutoff on the posterior probability of observing a meaningful treatment effect as we should have more evidence as the trial progresses; an arm stops for futility if this boundary is crossed. This futility stopping rule (equation \ref{eq:pip_fut2}) can be written in R as:
\begin{lstlisting}
futility2.arm.fun = (*@\textcolor{blue}{function}@*)((*@\textcolor{orange}{posterior, n, N}@*), (*@\textcolor{red}{b.fut, p.fut}@*)){
   (*@\textcolor{orange}{posterior}@*) < (*@\textcolor{red}{b.fut}@*)*(sum((*@\textcolor{orange}{n}@*))/(*@\textcolor{orange}{N}@*))^(*@\textcolor{red}{p.fut}@*)
}
\end{lstlisting}

An intervention may be declared effective at the end of the trial (at the maximum sample size) if the posterior probability of efficacy is above a certain (high) threshold:
\begin{equation}\label{eq:pip_eff}
P(\beta_{k} > \delta_e|\ \mathbf{y}, \mathbf{X}) \ \ > \ \ 1-b_e
\end{equation}

Here we use a value of $\delta_e = 0$ for the final look (here $\delta_e$ would also represent a clinically important treatment effect on the log-odds scale; this is the default value in \texttt{batss.glm()}). The value of $b_e$ is typically small and chosen to allow good control of type I error; we aim to control the pairwise one-sided type I error at $5\%$. The efficacy rule (equation \ref{eq:pip_eff}) can be written in R as:
\begin{lstlisting}
efficacy.arm.fun = (*@\textcolor{blue}{function}@*)((*@\textcolor{orange}{posterior}@*), (*@\textcolor{red}{b.eff}@*)){
  (*@\textcolor{orange}{posterior}@*) > 1-(*@\textcolor{red}{b.eff}@*) 
}
# where delta.eff needs to have NAs to not evaluate efficacy at interims
\end{lstlisting}


The trial will only stop early if all intervention arms have been dropped for futility; otherwise the trial will proceed to the maximum sample size of $N=216$ and evaluate efficacy and lack-of-benefit for the remaining arms at the end of the trial. Note that care should be taken when constructing the efficacy and futility stopping rules as it is possible to simultaneously meet the efficacy and futility criteria in certain situations (see Appendix C, Case Study 2.2).

Here we will explore two scenarios of interest, which are displayed in Table \ref{tab:pip_scenarios}. In the first scenario (null), all treatments have an ORR of 40$\%$. In the second (alternative) scenario, we have 1 minimally effective treatment and two highly effective treatments (along with two treatments with the same ORR as the control).

To assess the operating characteristics of our case study designs, we simulated 10,000 trials for each scenario. The design parameters for the user-defined functions (e.g., $b_e$, $b_f$, $p_f$, $\gamma$, $\eta$, $\nu$) were chosen to produce suitable operating characteristics (namely, pairwise one-sided type I error of 5$\%$ and 80-90$\%$ power to detect a large effect). The values for $p_f$, $\gamma$, $\eta$, $\nu$ were initially informed by previous works (\citealp{Cellamare2017}; \citealp{Gotmaker}) and then were tuned for our case study. $b_f$ (and to some extent, $b_e$) were also chosen based on clinicians' input (in addition to the operating characteristics).  The $\delta$ values were chosen based on clinicians' input to incorporate clinically important differences.

The pairwise one-sided type I error was calculated as the proportion of simulations that falsely declared efficacy when the intervention was assumed to have the same ORR as the control (i.e., log OR = 0). The (pairwise) power was calculated as the proportion of simulations that declared efficacy for an intervention when it was assumed to have a large effect (i.e., $>20\%$ absolute improvement in the ORR compared to the control; as in the ``alternative'' scenario).

\begin{table}
\Fontviii
    \centering
    \begin{tabular}{lcccccc}\hline
     \multicolumn{7}{l}{\textbf{6-month ORR}} \\ 
        \textbf{Scenario} & \textbf{Arm A (control)} & \textbf{Arm B}  & \textbf{Arm C} &  \textbf{Arm D} & \textbf{Arm E} & \textbf{Arm F} \\
        \cmidrule(r){1-1}
        \cmidrule(r){2-2}
        \cmidrule(r){3-3}
        \cmidrule(r){4-4}
        \cmidrule(r){5-5}
        \cmidrule(r){6-6}
        \cmidrule(r){7-7}        
         Global Null & 0.4 & 0.4 & 0.4 & 0.4 & 0.4 & 0.4 \\
         Alternative & 0.4 & 0.4 & 0.4 & 0.5 & 0.7 & 0.7 \\\hline
    \end{tabular}
    \caption{Scenarios explored for PIP trial case study (Case Study 2)}
    \label{tab:pip_scenarios}
\end{table}

Here we have focused on providing the user-defined functions for performing the adaptations. The full description of the \texttt{batss.glm()} call for simulating the ``Alternative'' scenario for Case Studies 2.1 and 2.2 can be found in Appendix C, following the logic described above.

\subsection{Results}\label{pipresults}

The output from the \texttt{summary} command can be found in Appendix C, along with an explanation of the output.

\subsubsection{Case Study 2.1}

\begin{figure}
   \centering
  \includegraphics[width=12.5cm]{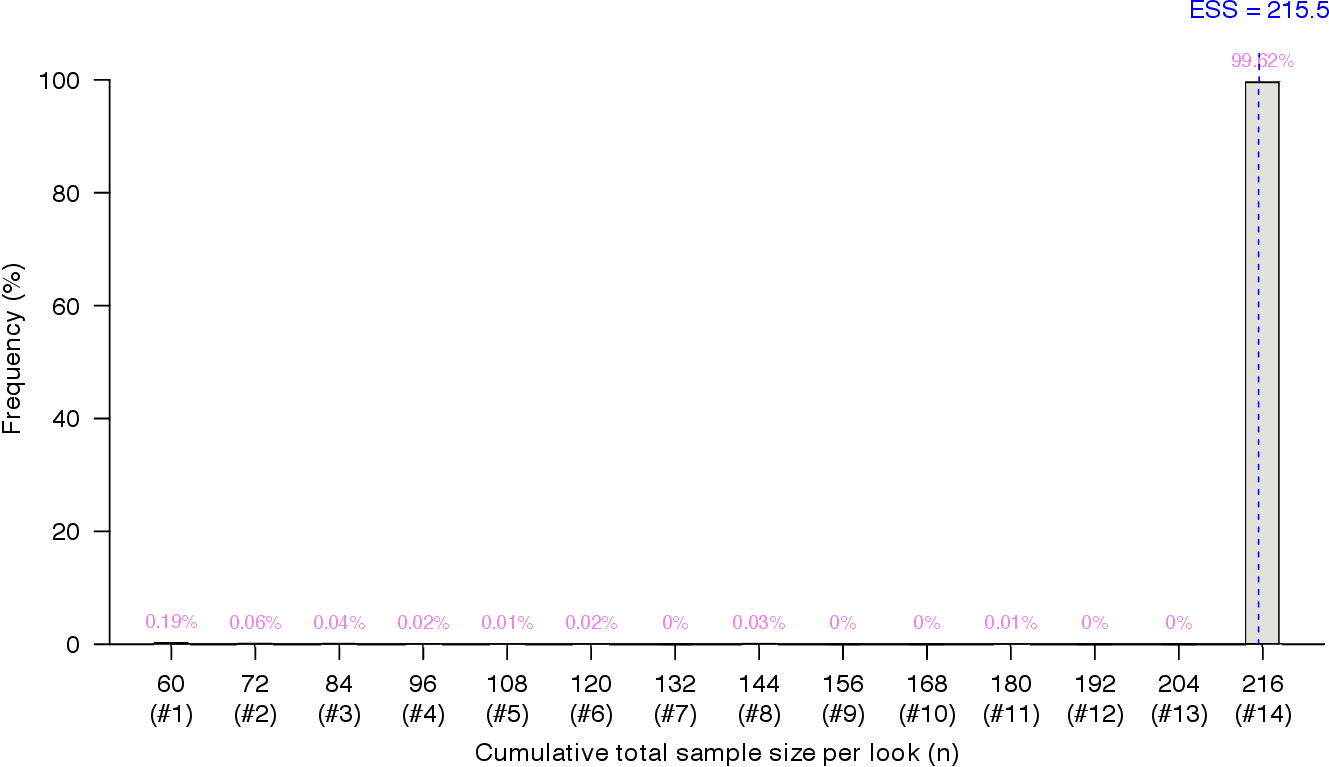}
 \caption{\Fontviii Barplot of the probability of stopping (y-axis) at each look (x-axis) as observed across the 10,000 simulated trials for the ``Alternative'' scenario for Case Study 2.1. The vertical blue dashed line shows the estimate of the expected sample size (ESS). As the trial only stops if all arms are dropped for futility, the expected sample size almost equals 216, the maximum sample size.}
    \label{fig:PIPplot2-e11}
\end{figure}

\begin{figure}
   \centering
  \includegraphics[width=\textwidth]{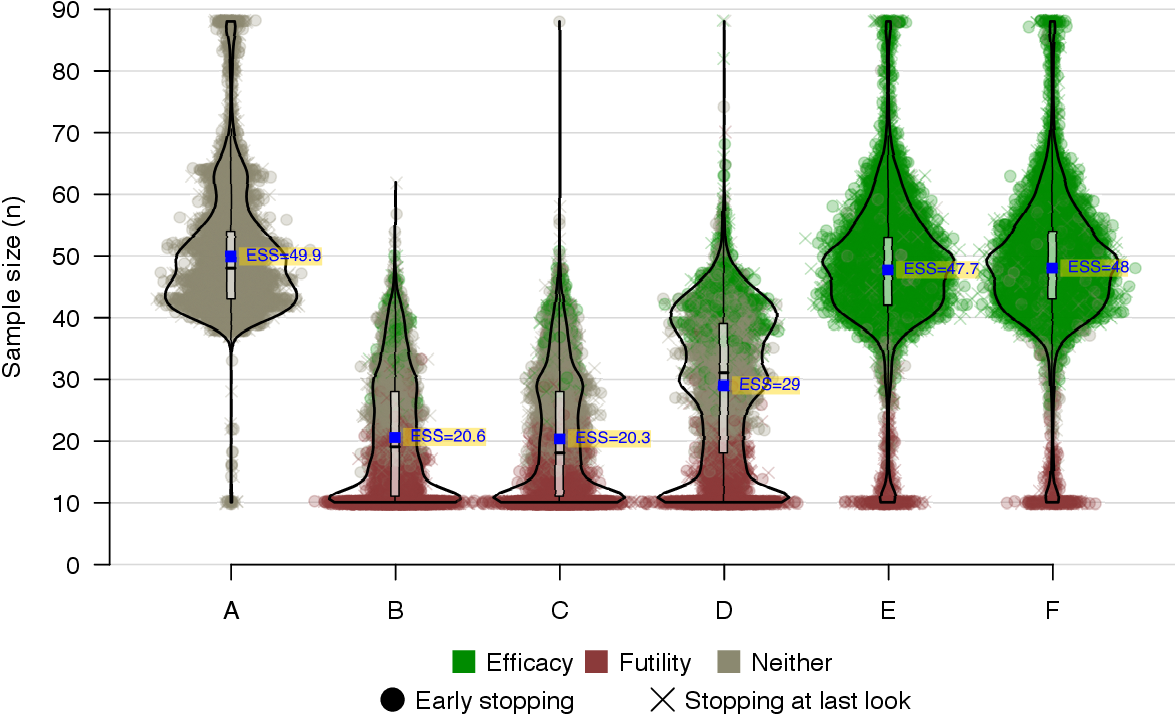}
    \caption{\Fontviii Violin plot (and boxplot) of the simulated sample sizes per group and corresponding expected sample sizes for the ``Alternative'' scenario for Case Study 2.1 across the 10,000 simulated trials. The sample size per group of each trial are displayed. For target parameters, these points are colour-coded and shaped by the decision and timing when the trial stopped. We can note that the ineffective arms `B' and `C' are are often dropped early for futility and that more patients are allocated to effective arms `E' and `F'.}
    \label{fig:PIPplot3-e11}
\end{figure}

\begin{figure}
   \centering
  \includegraphics[width=\textwidth]{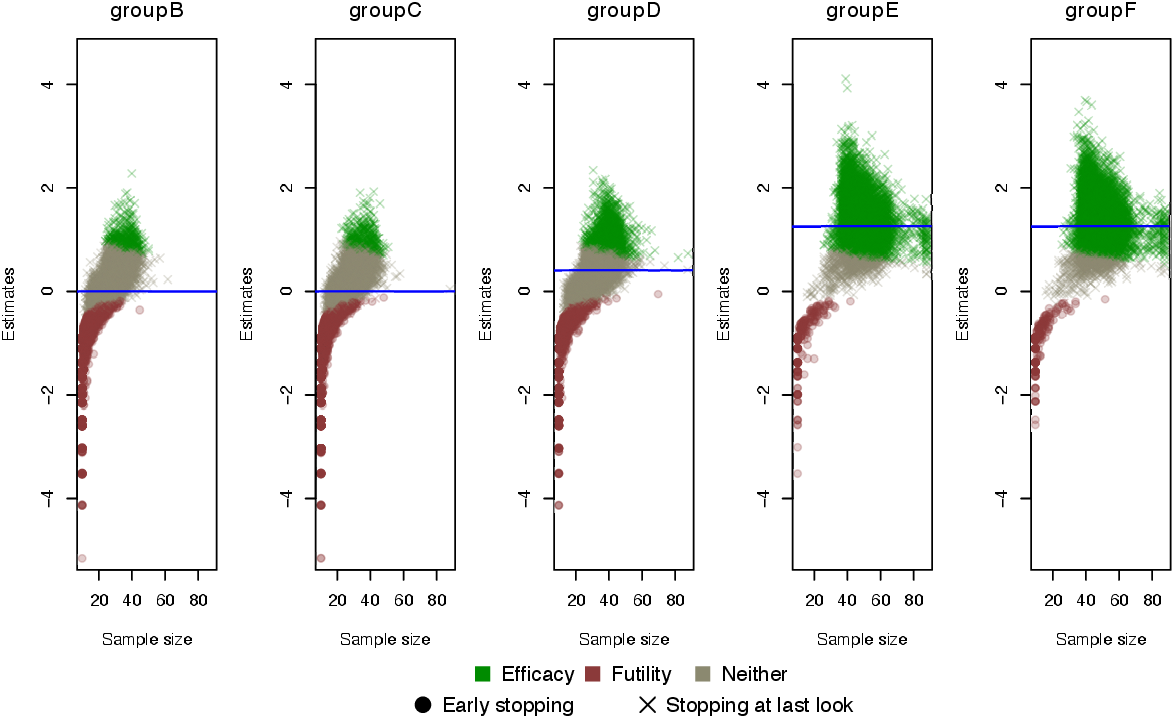}
 \caption{\Fontviii Scatterplot of target parameter log odds ratio estimates (posterior mean) versus sample size for each intervention arm (separate plots) at the time of stopping the arm, along with the decision at the time of stopping for the ``Alternative'' scenario for Case Study 2.1 across the 10,000 simulated trials. The timing of the stop is also provided (``Early stopping'' = stopping at an interim analysis). We can note that log odds ratio estimates are typically negative when dropping arms early for futility.}
    \label{fig:PIPplot4-e11}
\end{figure}

Table \ref{tab:pipresults11} shows the proportion of simulations that declared each intervention arm to be effective (at the maximum sample size/end of the trial), the proportion of simulations that each intervention arm was stopped early to declare futility, and the average sample size per arm and overall sample size, for each scenario for Case Study 2.1. Figure \ref{fig:PIPplot2-e11} shows the probability of stopping at a given look and confirms that early stopping -- only occurring when all arms are found futile -- is rare. The distribution of sample sizes for each treatment arm can also be examined in Figure \ref{fig:PIPplot3-e11} for the ``Alternative'' scenario. A summary of the target parameter estimates against the sample size achieved for a simulated trial is provided in Figure \ref{fig:PIPplot4-e11}.  

It can be seen that the pairwise 1-sided type I error is controlled at approximately 5$\%$ under the global null, and that there is approximately 90$\%$ power to declare efficacy for effective intervention arms (those with more than a 20$\%$ (absolute) ORR improvement). Intervention arms were stopped early 50-60$\%$ of the time for futility when they were assumed to have no benefit over the control, and early stopping for futility only occurred in a small proportion ($<5\%$) of simulations for effective intervention arms.

\begin{table}
\Fontviii
    \centering
    \begin{tabular}{lcccc}\hline
        & \multicolumn{2}{c}{\textbf{Conclusion}} & \multicolumn{2}{c}{\textbf{Sample size}}\\
        \cmidrule(r){2-3}
        \cmidrule(r){4-5}
        \textbf{Group} & \textbf{Efficacy$^{(1)}$} & \textbf{Futility$^{(2)}$}  & \textbf{Average$^{(3)}$}  &  \textbf{Stand. Dev.$^{(4)}$}  \\
        \cmidrule(r){1-1}
        \cmidrule(r){2-2}
        \cmidrule(r){3-3}
        \cmidrule(r){4-4}
        \cmidrule(r){5-5}
        \multicolumn{5}{c}{}\\
        \multicolumn{5}{l}{\textbf{Under the global null hypothesis:}}\\
Control (Arm A)&    -&    -& 49.04&17.29\\
\cmidrule(r){1-1}
Arm B&0.044&0.588& 28.37&17.98\\
Arm C&0.044&0.593& 27.87&17.77\\
Arm D&0.047&0.599& 27.54&17.66\\
Arm E&0.045&0.591& 28.12&17.86\\
Arm F&0.046&0.588& 28.10&17.95\\
\cmidrule(r){1-1}
At least 1&0.156&0.880&     -&    -\\
All&0.001&0.295&189.04&48.79\\
        \multicolumn{5}{c}{}\\
        \multicolumn{5}{l}{\textbf{Under the alternative hypothesis:}}\\
Control (Arm A)&    -&    -& 49.89& 9.49\\
\cmidrule(r){1-1}
Arm B&0.045&0.509& 20.58& 9.74\\
Arm C&0.041&0.508& 20.34& 9.67\\
Arm D&0.211&0.280& 28.95&12.20\\
Arm E&0.898&0.033& 47.71&10.65\\
Arm F&0.900&0.028& 48.01&10.46\\
\cmidrule(r){1-1}
At least 1&0.982&0.713&     -&    -\\
All&0.006&0.004&215.49& 8.54\\
\hline
    \end{tabular}
    \caption{$^{(1)}$ Probability of concluding a treatment arm, at least one treatment arm, or all treatment arms (arms B to F) to be  effective at the end of the trial, $^{(2)}$ probability of stopping early a treatment arm, at least one treatment arm, or all treatment arms for futility (arms B to F), $^{(3)(4)}$ expected sample size per arm (A to F) and in total, and corresponding standard deviation, under the null and alternative hypotheses (refer to Table \ref{tab:pip_scenarios}) in Case study 2.1 }
    \label{tab:pipresults11}
\end{table}



\subsubsection{Case Study 2.2}

Table \ref{tab:pipresults2} shows the results of the simulations using the \texttt{batss.glm} function for Case Study 2.2 (which used an increasing function for the futility boundaries). Again, the pairwise 1-sided type I error is controlled at  5$\%$ under the global null, but the power has decreased slightly using the increasing futility boundary values.

From Table \ref{tab:pipresults2} we can see that a higher proportion of early stopping for futility is occurring compared to Case Study 2.1 (which is to be expected), with over 90$\%$ of the simulations stopping early for futility for intervention arms that were assumed to have the same ORR as the control. A high proportion (79.9$\%$) of early stopping for futility also occurred for the intervention arm that was assumed to only have a 10$\%$ absolute improvement in the ORR (i.e., was a poorly-performing novel treatment). Compared to Case Study 2.1, the average sample sizes also decreased for poorly-performing arms.

\begin{table}
\Fontviii
    \centering
    \begin{tabular}{lcccc}\hline
        & \multicolumn{2}{c}{\textbf{Conclusion}} & \multicolumn{2}{c}{\textbf{Sample size}}\\
        \cmidrule(r){2-3}
        \cmidrule(r){4-5}
        \textbf{Group} & \textbf{Efficacy$^{(1)}$} & \textbf{Futility$^{(2)}$}  & \textbf{Average$^{(3)}$}  &  \textbf{Stand. Dev.$^{(4)}$}  \\
        \cmidrule(r){1-1}
        \cmidrule(r){2-2}
        \cmidrule(r){3-3}
        \cmidrule(r){4-4}
        \cmidrule(r){5-5}
        \multicolumn{5}{c}{}\\
        \multicolumn{5}{l}{\textbf{Under the global null hypothesis:}}\\
Control (Arm A)&    -&    -&  39.75&14.79 \\
\cmidrule(r){1-1}
Arm B&0.030&0.946& 22.98&13.56\\
Arm C&0.030&0.948& 22.70&13.55\\
Arm D&0.029&0.948& 22.42&13.45\\
Arm E&0.029&0.950& 22.89&13.58\\
Arm F&0.029&0.948& 22.90&13.73\\
\cmidrule(r){1-1}
At least 1&0.096&0.998&     -&    -\\
All&0.000&0.842&153.58&45.87\\
        \multicolumn{5}{c}{}\\
        \multicolumn{5}{l}{\textbf{Under the alternative hypothesis:}}\\

Control (Arm A)&    -&    -& 53.89&10.09\\
\cmidrule(r){1-1}
Arm B&0.041&0.931& 18.66& 8.65\\
Arm C&0.038&0.935& 18.48& 8.55\\
Arm D&0.187&0.748& 26.23&12.01\\
Arm E&0.863&0.115& 48.33&12.13\\
Arm F&0.873&0.106& 48.82&11.86\\
\cmidrule(r){1-1}
At least 1&0.968&0.985&     -&    -\\
All&0.006&0.025&214.38&12.52\\
\hline
    \end{tabular}
    \caption{$^{(1)}$ Probability of concluding a treatment arm, at least one treatment arm, or all treatment arms (arms B to F) are  effective at the end of the trial, $^{(2)}$ probability of stopping early a treatment arm, at least one treatment arm, or all treatment arms for futility (arms B to F), $^{(3)(4)}$ expected sample size per arm (A to F) and in total, and corresponding standard deviation, under the null and alternative hypotheses (refer to Table \ref{tab:pip_scenarios}) in Case study 2.2 }
    \label{tab:pipresults2}
\end{table}


\vspace{0.5cm}

Summary plots of the output for Case Study 2.2 can be found in Appendix C. The run time for these examples for a cluster, and standard Windows desktop are provided in the Appendix.


\section{Discussion}\label{discussion}

Here we have presented an R package, \texttt{BATSS} \citep{BATSS}, which provides a modular, flexible, structured framework to simulate Bayesian adaptive trials. \texttt{BATSS} is an open source, script-based package that is  extendable, and rich with options for adaptations for simulating Bayesian adaptive clinical trials. It allows for easy calculation of various trial operating characteristics and visualisation of some key metrics. We have provided an overview of the workflow and arguments required for the \texttt{batss.glm} function which can be used to simulate Bayesian adaptive trials whose outcome may be modelled by a GLM. 

The \texttt{BATSS} package will help fill the gap between methodology research and implementation of Bayesian adaptive designs in practice. The \texttt{BATSS} package was designed to be accessible to R users with varying levels of programming expertise, and it is hoped that this package will reduce time spent on programming by avoiding the
need to write simulation code from scratch for a range of Bayesian adaptive designs. 

We believe that \texttt{BATSS} offers a good balance of flexibility by allowing the user to specify their RAR functions and algorithms for treatment allocations, as well as their early stopping/arm dropping rules, whilst taking care of the
simulation work. This is an advantage over existing packages which may not offer sufficient flexibility in the scenarios and design settings that can be explored. A challenge of developing
trial simulation software is to make it generalisable enough without overwhelming new users with the required inputs. Thus, there is a
trade-off of flexibility and ease-of-use; to assist with this we provide several worked examples and functions for some commonly-used trial designs. Additionally, there are often implicit assumptions in the software code about the conduct of the clinical trial; we have tried to limit these assumptions by allowing user-specified functions and have described the assumptions
that the \texttt{batss.glm} function makes.

Extending the \texttt{BATSS} package is fairly easy. This version of \texttt{BATSS} allows for the simulation of Bayesian adaptive designs which may be modelled using GLMs. We anticipate that additional functions will be added to the \texttt{BATSS} R package over time, with different functions for different types of analyses and designs. Future work will seek to extend it so that Bayesian adaptive designs may be simulated for time-to-event outcomes.

\subsection{Limitations and Future Research}\label{limitations}

One of the main limitations with using R-INLA to perform the Bayesian approximations is that it is frequently being updated, which could potentially cause stability issues with the \texttt{BATSS} package (e.g., issues with implementing on cluster networks). Also, not all models are currently available in the INLA framework (e.g., heteroscedastic models) and so use of the \texttt{BATSS} package will be limited to the classes of models that are supported in the INLA framework (which appear to be sufficient for many commonly-used clinical trial applications). 

The \texttt{BATSS} package does not currently support platform trial designs where treatment arms may be added or those implementing multi-factorial designs (with multiple domains of treatment). This is an area for future research. We also plan to extend the package to
\begin{itemize}[itemsep=-5pt]
   \item support hierarchical models so that designs that implement information borrowing across subgroups can be implemented (e.g., basket trials). \cite{hosseini2023} provide evidence that this is feasible and fast using INLA, and may provide more computational savings compared to MCMC for these models.
    \item consider the enrollment times of patients or account for follow-up duration (leading to delays in data availability) or loss to follow-up and drop-out. Instead of assuming that analyses are conducted based on the number of participants completing follow-up (which, here, is also the same as the number recruited), a related extension would be to take into account the fact that some of the  participants recruited have incomplete follow-up at the time of an interim analysis. Such features, typically not included in existing software for simulating Bayesian adaptive designs -- with FACTS \citep{facts} being one of the few exceptions --, would allow estimating further operating characteristics, mentioned by the guidelines described in Section \ref{montecarlosim}, like, for example, the expected, minimum and maximum study times and number of recruited participants ($F^{19}_2$, $I^{20}_1$). We are currently extending \texttt{BATSS} to explicitly simulate accrual, drop-outs and account for the follow-up duration length (and therefore have participants with incomplete follow-up in the trial at interim analyses), and the next release of \texttt{BATSS} will contain these features. This is necessary in particular for survival/time-to-event endpoints.  
    \item propose point and interval estimators for the target parameters. Bias and mean squared error of point estimators as well as coverage of interval estimators are typically of interest for regulators ($F^{11}_2$, $I^{11}_1$, $I^{21}_1$). Whilst raw trial data for each simulated data set are available to users (by specifying \texttt{extended = 2} in \texttt{batss.glm}) to define the properties of their planned estimators (see, e.g., \cite{Robertson2023A}, \cite{Robertson2023B}, \cite{Robertson2025A}, \cite{Robertson2025B}), allowing this task to be performed with the same modular approach as to define other aspects of the trial, would be useful.
\end{itemize}
This version of \texttt{BATSS} is appropriate for slow recruiting trials and/or those that have endpoints that are observed over a short period (and no loss-to-follow-up). 

Other extensions include extending the \texttt{BATSS} package for proportional odds models; trial designs where there is no control or reference group or where all treatment arms will be compared to one another (to determine the ``best'' treatment). 

\section*{Data availability}
The datasets used in the manuscript, appendices, and online material (\url{https://batss-stable.github.io/BATSS/}) are synthetic and were produced by simulation, as described in Section \ref{section-batss}. No external or real-world datasets were used. For the purpose of open access, the author has applied a Creative Commons Attribution (CC BY) license to any Author Accepted Manuscript version arising.
\section*{Funding}
Dominique-Laurent Couturier and Prof. Thomas Jaki were supported by the Medical Research Council under Grant \verb+MC_UU_00040/03+. The funder had no role in study design,  analysis, decision to publish, or preparation of the manuscript.
\section*{Conflict of interest}
The authors declare that they have no competing interests.
\section*{Ethics approval and consent to participate}
This work is based on synthetic data and did not involve human participants, human material, or animals; therefore, ethical approval and consent to participate were not required.

\renewcommand\refname{References}
\Fontviii
  \bibliography{bibliography.bib}

\end{document}